\documentclass[12pt]{article}
\usepackage{graphicx}

\def\leaderfill{\leaders\hbox to 1em{\hss.\hss}\hfill}  %% òî÷êà-ïðîâîäíèê
\begin{document}

\noindent {\footnotesize\it  ISSN 1063-7737,
 Astronomy Letters, 2007, Vol. 33, No. 9, pp. 571--583. \copyright Pleiades Publishing, Inc., 2007.

\noindent Original Russian Text \copyright V.V. Bobylev, A.T.
Bajkova, 2007, published in Pis'ma v
Astronomicheski$\check{\imath}$ Zhurnal, 2007, Vol. 33, No. 9, pp.
643--656.}

%\vskip -4mm

\noindent
\begin{tabular}{llllllllllllllllllllllllllllllllllllllllllllll}
& & & & & & & & & & & & & & & & & & & & & & & & & & & & & & & & & & & & &  \\
\hline \hline
\end{tabular}

\vskip 1.5cm

 \centerline {\large\bf Kinematics of the Scorpius-Centaurus OB Association}
 \bigskip
 \centerline {V.V. Bobylev and A.T. Bajkova}
 \bigskip
 \centerline {\small\it
Central (Pulkovo) Astronomical Observatory of RAS, St-Petersburg,
}
 \bigskip

{\bf Abstract} A fine structure related to the kinematic
peculiarities of three components of the Scorpius–Centaurus
association (LCC, UCL, and US) has been revealed in the
$UV$-velocity distribution of Gould Belt stars. We have been able
to identify the most likely members of these groups by applying
the method of analyzing the two-dimensional probability density
function of stellar UV velocities that we developed. A kinematic
analysis of the identified structural components has shown that,
in general, the center--of--mass motion of the LCC, UCL, and US
groups follows the motion characteristic of the Gould Belt,
notably its expansion. The entire Scorpius-Centaurus complex is
shown to possess a proper expansion with an angular velocity
parameter of $46\pm8$ km s$^{-1}$ kpc$^{-1}$ for the kinematic
center with $l_0=-40^\circ$ and $R_0=110$ pc found. Based on this
velocity, we have estimated the characteristic expansion time of
the complex to be $21\pm4$ Myr. The proper rotation velocity of
the Scorpius–Centaurus complex is lower in magnitude, is
determined less reliably, and depends markedly on the data
quality.

\section*{INTRODUCTION}

When the first data on the spectral types of luminous stars became
available, numerous researchers identified well-defined groups of
O and B stars that subsequently, at the suggestion of Ambartsumian
(1947), came to be called associations. The hypothesis of a low
space density and, as a result, dynamical instability of
associations in the field of Galactic tidal forces was also first
put forward by Ambartsumian (1949).

As was shown by Blaauw (1952), an expanding association with an
initially spherical shape stretches into an ellipse whose
orientation changes with time as a result of the Galactic
differential rotation.

Scorpius-Centaurus (Sco OB2 or Sco-Cen) is the nearest
association. Blaauw (1964) was the first to estimate the kinematic
age of this association, $\approx$20 Myr, by analyzing stellar
radial velocities using a linear expansion coefficient $k=50$ km
s$^{-1}$ kpc$^{-1}$. This age corresponds to the time it takes for
a star to traverse the characteristic radius of the region
occupied by the association.

Following Blaauw (1964), one identifies three groups in the
Scorpius-Centaurus association: US (Upper Scorpius), UCL (Upper
Centaurus-Lupus), and LCC (Lower Centaurus-Crux). At present, many
authors suggest expanding the list of possible association
members. There is indubitable evidence that the recently
discovered nearby open clusters $\beta$ Pic, TWA, Tuc/Hor, $\eta$
Cha and $\epsilon$ Cha (Mamajek and Feigelson 2001; Song et al.
2003; Mamajek 2005) also belong to the Scorpius–Centaurus
association or are part of its loose halo.

The ages of the association members were estimated by de Geus et
al. (1989) using data in the Walraven (VBLUW) photometric system:
$5-6$ Myr for US, $14-15$ Myr for UCL, and $11-12$ Myr for LCC.
The current age estimates for the association members obtained by
comparing the evolutionary tracks of stars with their positions in
the Hertzsprung-Russell diagram (Mamajek et al. 2002; Sartori et
al. 2003) are $8-10$ Myr for US and $16-20$ Myr for UCL and LCC.

Based on their analysis of data from the Hipparcos Catalog (1997)
by the spaghetti method, de Zeeuw et al. (1999) compiled a list of
likely Scorpius-Centaurus members and determined their mean
distances: $145\pm2$ pc to US, $140\pm2$ pc to UCL, and $118\pm2$
pc to LCC.

A considerable number of low-mass late-type pre-main-sequence
stars with ages of several million years have been revealed in the
Scorpius-Centaurus association by several characteristic
signatures (lithium overabundance, X-ray emission, etc.) (Preibish
and Zinnecker 1999; Mamajek et al. 2002; Sartori et al. 2003).
There are both radial velocities and proper motions for some of
these stars (Makarov 2003; Mamajek et al. 2002; Mamajek 2005;
Torres et al. 2006).

In the opinion of Preibisch and Zinnecker (1999), the star
formation process in the Scorpius-Centaurus association finished
about a million years ago; it contains no significant (in mass)
reserves of dust, since its visual extinction is low
($A_v\leq2^m$), and of gas, which was swept up to the periphery of
the region after supernova explosions. This is an important
argument that the association is gravitationally unbound.

There are strong grounds for believing that the nearby OB
associations and, in particular, the Scorpius-Centaurus
association, are members of the Gould Belt (Lindblad et al. 1973;
de Zeeuw et al. 1999; Bobylev 2006).

A critical review of the formation scenarios for the
Scorpius-Centaurus association can be found in Sartori et al.
(2003). These authors considered the model of successive star
formation (Blaauw 1964, 1991; Preibisch and Zinnecker 1999), the
Gould Belt model (Lindblad et al. 1973; Olano 1982), and the model
of star formation by infall of high-velocity clouds on the
Galactic disk (L\'epine and Duvert 1994) and provided arguments
for the model related to the passage of a spiral arm near the Sun.

The goal of this paper is to refine the structure of the
Scorpius-Centaurus association and to determine its expansion and
rotation velocities from currently available observational data.

\section*{INPUT DATA}

Dravins et al. (1999), de Bruijne (1999), Lindegren et al. (2000),
and Madsen et al. (2002) showed that the individual distances for
stars of nearby open clusters, such as the Hyades, could be
determined by the well-known moving cluster method more accurately
than from Hipparcos parallaxes. The moving cluster method for
stars of the Scorpius-Centaurus association was used to determine
``improved'' individual distances (de Bruijne 1999) and ``improved
astrometric'' radial velocities (Madsen et al. 2002). Monte Carlo
simulations (de Bruijne 1999) showed that the
expansion/contraction of an association affects significantly the
radial velocity determinations and, to a lesser extent, the bias
of stellar distance estimators. The radial velocities determined
by Madsen et al. (2002) for a large number of stars in the
Scorpius-Centaurus association ($\sim$500 stars) from the list by
de Zeeuw et al. (1999) are of interest to us as a highly accurate
ephemeris.

There is a list of 700 Hipparcos stars with measured radial
velocities belonging to the Gould Belt at our disposal. Its
description can be found in our previous papers (Bobylev 2004b,
2006). For the next series of Scorpius-Centaurus stars, we made a
number of additions when new data became available:

(a) for the classical members of the Scorpius-Centaurus
association (UCL, LCC, and US), the list of candidates was taken
from de Zeeuw et al. (1999), which was expanded to include a
number of stars from the list by de Geus et al. (1989);

(b) for the $\beta$ Pic cluster, the new data were taken from
Torres et al. (2006);

(c) For the TWA cluster, the data were taken from Mamajek (2005);
in this paper, we use only five stars of this cluster with
measured trigonometric parallaxes---TWA 1, 4, 9, 11, and 19;

(d) the list of candidates of the Chamaeleon open cluster, which
also includes the small known $\eta$ Cha and $\epsilon$ Cha
clusters, was taken from Sartori et al. (2003).

The OSACA (Bobylev et al. 2006) and PCRV (Gontcharov 2006)
catalogs served as the main sources of stellar radial velocities.
The main peculiarity of these catalogs is that they were reduced
to the same radial velocity standard proposed by Gontcharov. The
difference between these two versions is that the OSACA catalog
includes the ``astrometric'' radial velocities of
Scorpius-Centaurus stars from Madsen et al. (2002), while these
were not used in compiling the PCRV catalog.

For several members of the Scorpius-Centaurus association (UCL,
LCC, and US), we calculated new stellar radial velocities based on
the PCRV catalog (Gontcharov 2006) and the paper by Jilinski et
al. (2006). No stars with a random error in the radial velocity
larger than 5 km s$^{-1}$ were used. We designate the stellar
space velocities calculated using the radial velocities from these
catalogs as ``real'' to distinguish them from those calculated
using the ``astrometric'' radial velocities from Madsen et al.
(2002).

As the stellar distance estimates, we use only the Hipparcos
trigonometric parallaxes. We use only those stars that satisfy the
condition $e_\pi/\pi<0.2$, which allows the influence of the
well-known Lutz-Kelker effect (Lutz and Kelker, 1973) to be
reduced. We took into account the Galactic rotation using the Oort
constants
 $A= 13.7\pm0.6$ km s$^{-1}$ kpc$^{-1}$ and
 $B=-12.9\pm0.4$ km s$^{-1}$ kpc$^{-1}$ determined previously (Bobylev 2004a).

Figure 1 shows the distribution of $UV$ velocities corrected for
the Galactic rotation for the stars that satisfy the condition
 $e_\pi/\pi<0.2$. Figure 1a presents  the $UV$ velocities of 255 Gould Belt
stars; only the ``real'' data were used to calculate the space
velocities. Figure 1b presents the UV velocities of 380 Scorpius–
Centaurus stars whose radial velocities were taken from the
catalog by Madsen et al. (2002). As can be seen from Fig.~1b,
three groups of stars related to the kinematic peculiarities of
UCL, LCC, and US are clearly identified. In Fig.~1a, this
separation into three isolated fractions is less distinct. Below,
we set the goal of separating the Scorpius-Centaurus structure
from the overall distribution of $UV$ velocities for the Gould
Belt based on the ``real'' data.

\section*{METHODS AND APPROACHES}
\subsection*{The Kinematic Model}

In this paper, we use a rectangular Galactic coordinate system
with the axes directed away from the observer toward the Galactic
center ($l=0^\circ$, $b=0^\circ$, the $x$ axis), along the
Galactic rotation ($l=90^\circ$, $b=0^\circ$, the $y$ axis), and
toward the North Galactic Pole ($b=90^\circ$, the $z$ axis).

We apply the equations derived from Bottlinger's standard formulas
(Ogorodnikov 1965) by assuming the existence of a common kinematic
center for rotation and expansion using two terms of the Taylor
expansion of the angular velocity for rotation $\omega$ and the
analogous parameter for expansion $k$ (Lindblad 2000; Bobylev
2004b). The conditional equations are
$$
\displaylines{\hfill
  V_r= -U_0\cos b\cos l-\hfill\llap{(1)}
\cr\hfill
      -V_0\cos b\sin l
      -W_0\sin b+\cos^2 b k_0 r+
\hfill\cr\hfill
  +(R-R_0)(r\cos b-R_0\cos (l-l_0))\cos b  k'_0-
\hfill\cr\hfill
 -R_0 (R-R_0)\sin (l-l_0) \cos b
 \omega'_0,\hfill\cr
\hfill V_l= U_0\sin l
           -V_0\cos l - \hfill\llap(2)\cr\hfill
  -(R-R_0)(R_0\cos (l-l_0)-r\cos b)\omega'_0+
\hfill\cr\hfill
 +r\cos b \omega_0+R_0(R-R_0)\sin (l-l_0) k'_0,\hfill
 \cr
\hfill V_b= U_0\cos l\sin b+ \hfill\llap(3) \cr\hfill
           +V_0\sin l \sin b
           -W_0\cos b-\cos b\sin b k_0 r-
\hfill\cr\hfill
  -(R-R_0)(r\cos b-R_0\cos (l-l_0))\sin b k'_0+
\hfill\cr\hfill
 +R_0(R-R_0)\sin (l-l_0)\sin b\omega'_0.\hfill
 }
$$
Here,
 $V_r$ is the stellar radial velocity,
 $V_l= 4.74r \mu_l \cos b$,
 $V_b= 4.74r \mu_b$, the coefficient 4.74 is the quotient of the number of
kilometers in an astronomical unit by the number of seconds in a
tropical year, $r=1/\pi$ is the heliocentric distance of the star,
the stellar proper motion components $\mu_l \cos b$ and $\mu_b$
are in mas yr$^{-1}$, the radial velocity
 $V_r$ is in km s$^{-1}$, the
parallax $\pi$ is in mas,
 $U_0$, $V_0$, $W_0$
are the stellar centroid velocity components relative to the Sun,
$R_0$ is the distance from the Sun to the kinematic center, $R$ is
the distance from the star to the center of rotation, $l_0$ is the
direction of the kinematic center, $R$, $R_0$, and $r$ are in kpc.
The quantity $\omega_0$ is the angular velocity of rotation and
$k_0$ is the radial expansion/contraction velocity of the stellar
system at distance $R_0$; the parameters $\omega'_0$ and $k'_0$
are the corresponding derivatives. The distance $R$ can be
calculated using the expression
$$
\displaylines{\hfill
 R^2=(r\cos b)^2-2R_0 r\cos b\cos (l-l_0)+R^2_0.\hfill
 }
$$
The system of conditional equations (1)--(3) contains seven
unknowns: $U_0$, $V_0$, $W_0$ $\omega_0$, $\omega'_0$,
 $k_0$ and $k'_0$, to be determined by
the least-squares method. Equations (1)--(3) are written in such a
way that the rotation from the $x$ axis to the $y$ axis is
considered positive.

\subsection*{The Method for Separating Stellar Fractions}

In this paper, we develop a variety of the probabilistic approach
to separating stellar fractions based on the approximation of the
two-dimensional probability density function of the $UV$
velocities of all the stars under consideration by a set of
individual Gaussians that represent the probability densities of
individual structural features.

To estimate the two-dimensional probability density $f(U,V)$ for
all the stars under consideration, we apply an adaptive kernel
method (Skuljan et al. 1999) to the map of initial, discretely
distributed space UV velocities. In contrast to Skuljan et al.
(1999), we use a two-dimensional, radially symmetric Gaussian
kernel function:
$$
\displaylines{\hfill
 K(r)= {1\over\sqrt{2\pi\sigma}}\exp-{r^2\over2\sigma^2},\hfill
 }
$$
where $r^2 = x^2 + y^2$. Obviously, the relation $\int K(r)dr=1$
needed to estimate the probability density holds. Note that, in
our case, the bin size of the two-dimensional maps was chosen to
be 0.25 km s$^{-1}$, given the density of star distribution in the
velocity field; the area of a square bin was $S=0.25\times0.25$
km$^2$ s$^{-2}$.

The main idea of the adaptive kernel method is that a convolution
with a kernel whose width is specified by a parameter $\sigma$
that changes with the data density near a given point is performed
at each point of the map. Thus, in zones with an enhanced density,
the smoothing is performed with a relatively narrow kernel.

We use the following definition of the adaptive kernel estimator
at an arbitrary point $\xi=(U,V)$ (Skuljan et al. 1999):
$$
\displaylines{\hfill
 \hat{f}(\xi)={1\over n}\sum_{i=1}^n K\Biggl({{\xi-\xi_i}\over h\Lambda_i}\Biggr),\hfill
 }
$$
where $\xi_i=(U_i,V_i)$, $\Lambda_i$ is the local dimensionless
bandwidth factor at point $\xi_i$, $h$ is a general smoothing
parameter, and $n$ is the number of data points $\xi_i=(U_i,V_i)$.
The local bandwidth factor $\Lambda_i$ at each point of the
two-dimensional $UV$ plane is defined by
$$
\displaylines{\hfill
 \Lambda_i= \sqrt{g\over \hat{f}(\xi)},\hfill\cr\hfill
    \log g={1\over n}\sum_{i=1}^n \ln \hat{f}(\xi),\hfill
 }
$$
where $g$ is the geometric mean of $\hat{f}(\xi)$. Obviously, to
compute $\Lambda_i$, we need the distribution estimate $f(\xi)$,
which, in turn, can be computed only when all $\Lambda_i$ are
known. Therefore, the problem of finding the sought for
distribution can be solved iteratively. As the first
approximation, we use the distribution obtained by smoothing the
initial $U$V map with a fixed kernel. An optimal value for the
smoothing parameter $h$ is determined by minimizing the rms
deviation of the $\hat{f}(\xi)$ estimate from the true
distribution $f(\xi)$. The value of $h$ that we found for the
complex of stars considered here is about 1.25 km s$^{-1}$. In
addition, the maps are scaled by the factor $n\cdot S$ to obtain
the total probability equal to unity.

Our two-dimensional probability density map of stellar $UV$
velocities (derived from the data of Fig.~1a) is presented in
Fig.~2a. In this figure, we can visually identify the
distributions of LCC, UCL, and US stars and an extended component
of the Gould Belt. Although the distributions of LCC, UCL, and US
stars in Fig.~2a are less isolated than those in Fig.~1b, in
general, we can see a similarity in structure between the $UV$
velocity planes constructed using fundamentally different (with
regard to the stellar radial velocities) data.

With the goal of a formally justified assignment of each star from
the complex under consideration to a particular fraction, we
approximate each fraction by a set of Gaussians, each of which is
represented by six unknown parameters:
$$
\displaylines{\hfill
 P_i(U,V)= A_i \exp[ -{{((U-U_{0i})\cos\theta_i+(V-V_{0i})\sin\theta_i))^2}\over{2\sigma^2_{Ui}}}~-\hfill\cr\hfill
 -{{((U-U_{0i})\sin\theta_i-(V-V_{0i})\cos\theta_i))^2}\over{2\sigma^2_{Vi}}}
  \Biggr],\hfill
 }
$$
where $i$ is the Gaussian number, $P_i(U,V)$ is the corresponding
distribution function, $A_i$ is the amplitude proportional to the
number of stars in the fraction,
 $U_{0i}$ and
 $V_{0i}$ are the $U$ and $V$
coordinates of the Gaussian center, respectively,
 $\sigma_{Ui}$ and
 $\sigma_{Vi}$ are
the corresponding dispersions of the distribution, and $\theta_i$
is the angle defining the Gaussian orientation in the $UV$ plane
(the angle between the major axis of the ellipse and the $V$
axis).

In the general case, we represent the LCC fraction by the sum of
 $N^{LCC}$ Gaussians, the UCL fraction by the sum of
 $N^{UCL}$ Gaussians, the US fraction by the sum of
 $N^{US}$  Gaussians, and,
finally, the overall extended parts of the Gould Belt referring to
the entire complex of stars by the sum of
 $N^{Gould}$ Gaussians. Denote
the set of numbers of the Gaussians approximating the LCC, UCL, US
fractions and the extended parts of the Gould Belt by
 $N_{LCC}$, $N_{UCL}$, $N_{US}$, and $N_{Gould}$, respectively.
Next, we add all Gaussian $N=N^{UCL}+N^{UCL}+N^{US}+N^{Gould}$ to
obtain the model distribution
$$
\displaylines{\hfill
 P_M(U,V)= \sum_{i=1}^{N} P_i(U,V),\hfill
 }
$$
with $6\times N$ unknown parameters. To determine them, we apply
the least-squares method to bring the model distribution closer to
that obtained from the initial stellar velocity distribution
(Fig.~2a). Thus, we minimize the functional
$$
\displaylines{\hfill
 e=\sum_{U,V} \Vert f(U,V)-P_M(U,V)\Vert^2.\hfill
 }
$$
Next, we calculate the probabilities that each star with number
$k=1,2,...,n$ belongs to ones of the LCC, UCL, and US structures
or the extended part of the Gould Belt using the formulas
$$
\displaylines {\hfill
 P_k      = \sum_{i=1}^N P_i(U_k,V_k),\hfill\llap(4)\cr\hfill
 P_k^{LCC}= (\sum_{i\in N_{LCC}}P_i(U_k,V_k))/P_k,\hfill\cr\hfill
 P_k^{UCL}= (\sum_{i\in N_{UCL}}P_i(U_k,V_k))/P_k,\hfill\cr\hfill
 P_k^{US} = (\sum_{i\in N_{US}}P_i(U_k,V_k))/P_k,\hfill\cr\hfill
 P_k^{Gould}= (\sum_{i\in N_{Gould}} P_i(U_k,V_k))/P_k.\hfill
 }
$$
whose name is given as the superscript in the above formulas. In
our case, having solved the minimization problem at
 $N_{LCC}=1$,
 $N_{UCL}=2$,
 $N_{US} =1$, and
 $N_{Gould}=1$, we found a model distribution shown
in Fig.~2b. The approximation error is $4.4\%$. A higher accuracy
can be achieved by including more Gaussians approximating even
finer structural features in the $UV$ distribution.

\section*{RESULTS}

The columns of Table 1 give the following: (1) the Hipparcos
(1997) star number; (2) the cluster name, the letter g for US,
UCL, and LCC marks the stars from the list by de Geus (1989); (3)
the radial velocity; (4) the random error in the radial velocity;
(5) the number of catalogs used to calculate the mean radial
velocity; (6) a reference to the radial velocity source; (7)--(10)
the probabilities $P^{LCC}, P^{UCL}, P^{US}$, and $P^{Gould}$ that
we calculated using Eqs.~(4).

Table 1 is fully accessible in electronic form. In the electronic
version of the table, columns 11--20 give the parallaxes,
coordinates, proper motions, and photometric magnitudes copied
from the Hipparcos Catalog (1997) and the stellar space velocities
$U,V,W$ that we calculated. In all, it contains data on 255 stars
satisfying the condition $e_\pi/\pi<0.2$ that served as a basis
for constructing Figs.~1a and 2.

Practice has shown that the condition $P^{Gould}<1.0$ effectively
separates the background stars (in our case, those of the Gould
Belt) from the overall distribution of Scorpius-Centaurus stars.
We use this condition below.

To reduce the influence of the Lutz-Kelker effect (Lutz and Kelker
1973), below we use only the stars that satisfy the condition
$e_\pi/\pi<0.15$.

\subsection*{The Hertzsprung-Russell Diagram}

Figure~3 shows the color--absolute magnitude diagram for 134
stars, $e_\pi/\pi<0.15$. The isochrones for three ages constructed
from the data by Schaller et al. (1992) for metallicity $Z=0.02$
are shown. To take into account the interstellar extinction, we
used the $A_v$ estimates from Sartori et al. (2003) and de Bruijne
(1999). To pass from the Tycho photometric magnitudes $V_T$ and
$B_T$ to the Johnson system and to determine $V$ and $B-V$ , we
used the polynomials from Mamajek et al. (2002). In accordance
with the constructed diagram, we excluded several stars in the
list by de Zeeuw et al. (1999) belonging to the horizontal giant
branch from the list of candidates following the recommendations
by Sartori et al. (2003) and Mamajek et al. (2002).

\subsection*{The Spatial Structure of the Association}

Given the satisfaction of all the above conditions and
restrictions, apart from the classical LCC, UCL, and US members (a
total of 90 stars), the association of 134 stars includes 44 stars
belonging to the following clusters: $\beta$ Pic ($20\pm10$ Myr;
Barrado y Navascu\'es et al. 1999)---16 stars; nearby X-ray stars
from the lists by Makarov (2003)---9 stars and Wichmann et al.
(2003)---1 star (HIP 58996); TWA ($\approx$10 Myr; Mamajek et al.
2000)---5 stars; Tuc/Hor ($\approx$30 Myr; Torres et al. 2000;
Zuckerman et al. 2001)---5 stars; $\alpha$ Car---4 stars; Cha---3
stars; and IC~2602---1 star (HIP 52370).

We believe that one star of the compact open cluster IC~2602
($\approx$30 Myr; Luhman 2001) could have been included in the
sample by chance due to the dispersion of errors in the space
velocities.

Table~1 includes eight stars of the Chamaeleon cluster, most of
which have $P^{Gould}$ close to unity. Only for two stars, HIP
42637 ($\eta$ Cha, 8 Myr; Mamajek et al. 1999) and HIP 58484
($\epsilon$ Cha, $5-15$ Myr; Mamajek et al. 2000), did we obtain
noticeable values of $P^{LCC}\approx P^{UCL}\approx0.2$. We may
conclude that the expansion of the stellar composition of the
Chamaeleon cluster suggested by Sartori et al. (2003) was
performed by including stars from the loose halo of the
Scorpius-Centaurus association and stars belonging to the Gould
Belt structure.

The fact that we obtained high probabilities,
 $P^{LCC}\approx0.5$
or
 $P^{UCL}\approx0.5$, for several stars of the relatively old open
cluster $\alpha$ Car (``Platais 8'', 56~Myr; Piskunov et al. 2006)
is of considerable interest. Our list of stars belonging to the
Scorpius-Centaurus association agrees with the list by Fern\'andez
et al. (2006). Based on the epicyclic approximation, these authors
showed that 10--20 Myr ago, the separation between the LCC center
and such clusters as $\beta$ Pic, TWA, Tuc/Hor, $\eta$ Cha, and
$\epsilon$ Cha was at a minimum, being several tens of parsecs.

The total number of ``real'' stars is approximately equal to the
number of stars with available high precision radial velocities
(the database has not yet been published) in the list by Fuchs et
al. (2006) (79 classical members of the Scorpius-Centaurus
association).

\subsection*{The Expansion and Rotation of the Association}

Table~2 lists the kinematic parameters that we found from the
solution of Eqs.~(1)--(3) using only those Scorpius-Centaurus
stars for which the condition $e_\pi/\pi<0.15$ was satisfied. The
$U_0, V0, W0$ coordinates listed in the table are given with the
opposite sign for the convenience of their comparison with the
data in Figs.~1 and 2.

The parameters in the upper part of the table (solutions nos. 1,
2, 3) were found from the ``real'' data. The parameters in the
lower part of the table (solutions nos. 2a and 2b) were found from
the data by Madsen et al. (2002). Solution no.~2b was obtained
without using stellar radial velocities.from the simultaneous
solution of only two equations, (2) and (3). As can be seen from
the table, there are no significant differences between solutions
nos.~2a and 2b.

Solution no.~1 was obtained from the stellar velocities without
any corrections being applied.

Solutions no.~2 were obtained from the stellar velocities
corrected for the differential rotation of the Galaxy.

For solution no.~3, the stellar velocities were corrected both for
the Galactic rotation and for the motion (rotation and expansion)
of the Gould Belt. For this purpose, we place the kinematic center
of the Gould Belt at the point with $l_0=128^\circ$ and $R_0=150$
pc (Bobylev 2004b) and use
 $\omega_0=-19\pm5$ km s$^{-1}$ kpc$^{-1}$,
 $\omega'_0=33\pm16$ km s$^{-1}$ kpc$^{-2}$,
 $k_0 = 22\pm 5$ km s$^{-1}$ kpc$^{-1}$, and
 $k'_0=-58\pm16$ km s$^{-1}$ kpc$^{-2}$
found by analyzing the motions of the open clusters belonging to
the Gould Belt structure (Bobylev 2006).

In Fig. 4, the expansion, $V_R$, and rotation, $V_\theta$,
velocities of the selected 134 stars are plotted against distance
R from the kinematic center of the Gould Belt calculated at $l_0 =
128^\circ$ and $R_0=150$ pc. The velocities were reduced to the
center of mass of the Gould Belt. The plots show the expansion and
rotation curves for the Gould Belt that we found from the data on
open clusters (Bobylev 2006).

The parameters of the kinematic center $l_0$ and $R_0$ must be
known to determine the proper rotation and expansion parameters
for the association from Eqs.~(1)--(3). As the first
approximation, we assume that the direction of the kinematic
center of the association is close to that of its geometrical
center $l_0\approx45$, which we then refine. The results of
solving Eqs.~(1)--(3) obtained at fixed $l_0=45^\circ$. and at
various $R$ are presented in Fig.~5. As can be seen from the
figure, the derivative of the expansion velocity $k'_0$ is equal
to zero at $R_0=110$ pc. Similarly, having fixed $R_0$ and varying
$l_0$, we found a new approximation, $l_0=-40^\circ$.

Figure 6 presents the space velocities of 134 stars projected onto
the radial (away from the association center) and tangential
directions. The velocities were decomposed with the parameters of
the center $l_0=-40^\circ$ and $R_0=110$ pc. The velocities were
reduced to the center of mass of the association (based on the
data of Table~2). The plots show the straight lines that were
constructed with $k_0=46$ km s$^{-1}$ kpc$^{-1}$ for the linear
expansion coefficient and $\omega_0=29$ km s$^{-1}$ kpc$^{-1}$ for
the angular velocity of rotation found in solution no.~2 of
Table~2.

Thus, having a fairly complex model (Eqs.~(1)--(3)), we used the
derivatives $k'_0$ and $\omega'_0$ found, which are determined
with large errors, to improve the parameters of the kinematic
center and, in the long run, reduce the problem to the linear
case. As can be seen from Fig.~6a, the data are described well by
the linear expansion coefficient. The rotation is of lesser
interest to us, because $\omega_0$ (Fig.~5) is almost always
smaller and less stable than $k_0$.

Figure~7 is similar to Fig.~4, but here we use the velocities of
489 ($e_\pi/\pi<0.5$) Scorpius-Centaurus stars (only the classical
members of the association from the list by de Zeeuw et al.
(1999), whose radial velocities were taken from the catalog by
Madsen et al. (2002). The main difference between Figs.~7 and 4 is
related to the velocity $V_\theta$. As can be seen from Fig.~7b, a
distinct and fairly symmetric crowding of stars is observed at
 $R<300$ pc; just as in Fig.~4, the center of this distribution
deviates noticeably from the rotation curve of the Gould Belt. At
 $R>300$ pc, the distribution of stars has a different pattern.they
fall nicely on the rotation curve of the Gould Belt. In our
opinion, an appreciable fraction of the stars from the list by de
Zeeuw et al. (1999) do not belong to the Scorpius-Centaurus
association; they fell into the sample by chance. It may also be
concluded that the negative sign of $\omega_0$ (solutions nos. 2a
and 2b in Table~2) that we found from the data by Madsen et al.
(2002) does not reflect the essence of the real rotation of the
association, but is related to the peculiarity of the sample of
these stars.

The data obtained lead us to conclude that the Scorpius-Centaurus
complex is involved in the systematic expansion of the Gould Belt.
On the other hand, there is a proper expansion (solution no.~2 in
Table~2) with an angular velocity $k_0=46\pm8$ km s$^{-1}$
kpc$^{-1}$. Using the velocity $k_0$ obtained, we can estimate the
characteristic expansion time of the complex from the well-known
formula $T=977.5/k_0$ (Murray 1986), which is $T=21\pm4$ Myr and
is an independent kinematic estimate of the age for the
Scorpius-Centaurus complex. This value is in good agreement with
the estimate of $T=16-20$ Myr for the age of the oldest LCC and
UCL groups obtained by Sartori et al. (2003) from the analysis of
photometric data.

Based on the velocity $k_0 = 39\pm12$ km s$^{-1}$ kpc$^{-1}$
(solution no. 2b in Table~2) found without using stellar radial
velocities, we obtain a longer expansion time of the complex,
$T=25\pm8$ Myr, which agrees, within the $1\sigma$ error limits,
with the value obtained from the three observed velocity
components.

\section*{DISCUSSION}

We obtained contradictory data on the direction of rotation of the
association. Analysis of the motion of the Scorpius-Centaurus
association in the field of attraction of the Galaxy modeled in
the epicyclic approximation by Breitschwerdt and Avillez (2006)
and Fuchs et al. (2006) leads us to conclude that the observed
rotation most likely must have the positive sign. On this basis,
we conclude that the results obtained using the ``real'' data are
preferred.

The linear expansion coefficient that we found for the
Scorpius-Centaurus association, $k_0=46\pm8$ km s$^{-1}$
kpc$^{-1}$, is in good agreement with $k_0=50$ km s$^{-1}$
kpc$^{-1}$ obtained by Blaauw (1964) from the analysis of stellar
radial velocities. Having analyzed the radial velocities of 19
most likely members of the TWA cluster by Blaauw's method, Mamajek
(2005) found $k_0=49\pm27$ km s$^{-1}$ kpc$^{-1}$. For the members
of the moving cluster $\beta$ Pic (about 40 stars), Torres et al.
(2006) found a dependence of the velocity $U$ on the $x$
coordinate with a coefficient of 53 km s$^{-1}$ kpc$^{-1}$, which
is interpreted as the expansion parameter of this cluster.

A whole system of cold molecular clouds such as R CrA, Lupus,
Chamaeleon-Muska, and Coalsack, is known in a wide neighborhood of
the Scorpius-Centaurus association (Corradi et al. 2004). The
analysis by Corradi et al. (2003) revealed an outflow of gas (a
component located at a distance $d\leq60$ pc) from the
Scorpius-Centaurus association toward the Sun with a velocity of
$-7$ km s$^{-1}$. This is in good agreement with the expansion
model of the Scorpius-Centaurus complex.

Fern\'andez et al. (2006) showed that the onset of star formation
in the Scorpius-Centaurus complex could be related to the passage
of a giant molecular cloud through a spiral arm $\approx$30 Myr
ago. The authors believe that the model requires no contraction of
the entire cloud and that the onset of star formation in a small
region of the cloud with the highest pressure served as a
triggering mechanism. In the coordinate system associated with the
Local Standard of Rest, the parent cloud was in the first Galactic
quadrant and had coordinates $x\approx100$ pc and $y\approx100$ pc
(the directions are indicated in our coordinate system). This
model depends strongly on the assumed pattern speed, which is
presently known with an insufficient accuracy (Mel'nik et al.
2001; Popova and Loktin 2005).

In the opinion of Breitschwerdt and Avillez (2006), supernova
explosions in the Scorpius-Centaurus association are responsible
for the presence of two regions of hot coronal gas---the Local and
Loop I Bubbles. It was concluded that the number of supernova
explosions needed for the formation of the Local Bubble was
14--20.

Based on the epicyclic approximation, Fuchs et al. (2006) showed
that the separation between the LCC, UCL, and US centers and the
Local Bubble was at a minimum $10-15$ Myr ago.

Comparison of the $UV$-distributions for 302 Hipparcos stars
associated with the Scorpius-Centaurus association presented in
Fig.~3 from Fuchs et al. (2006) and Figs. 1 and 2 in this paper
shows a similarity in that the distribution ellipse is oriented at
an angle of about $45^\circ$, which is attributable to the
expansion of the complex.

The theory of successive star formation (Blaauw 1964, 1991) runs
into the difficulty that the Scorpius-Centaurus association
exhibits no noticeable gradient in stellar ages as a function of
the distance (the distance from the supernova). To eliminate this
peculiarity, Preibisch and Zinnecker (1999) assumed for US that
the stars were formed during a short time interval. The structure
of the $UV$-velocity distribution consisting of three fairly
isolated, approximately parallel branches that we found leads us
to conclude that there were three periods of star formation
separated by short time intervals in the expanding
Scorpius-Centaurus complex. We believe that this conclusion is
consistent with the conclusions by Preibisch and Zinnecker (1999).

\section*{CONCLUSIONS}

We considered the list of Hipparcos stars belonging to the
Scorpius-Centaurus association with measured parallaxes, proper
motions, and radial velocities.

Three characteristic features related to the kinematic
peculiarities of the LCC, UCL, and US groups were revealed in the
$UV$-velocity distribution of Gould Belt stars. A method for
analyzing the UV plane was developed to separate them from the
Gould Belt stars. Applying this method allowed us to identify the
most likely members of these features.

We believe the results obtained using the ``real'' data to be most
reliable. The stellar radial velocities calculated by the moving
cluster method from Madsen et al. (2002) have small random errors.
However, the list of candidates for Scorpius-Centaurus members by
de Zeeuw et al. (1999) probably contains a considerable number of
background stars (notably for distant stars farther than 150 pc
from the Sun) and needs to be corrected. We used a method for
separating the association members from the Gould Belt
distribution that is free from any initial assumptions about the
membership of stars in this association to a much larger degree
than the moving cluster method. At the same time, our method is
largely determined by the accuracy of the observational data and,
in particular, by the random errors in the stellar radial
velocities.

Analysis of the identified stars leads us to conclude that, in
general, the motion of the centers of mass of the three groups,
LCC, UCL, and US, follows the motion characteristic of the Gould
Belt; more specifically, they are involved in its rotation and, in
particular, in its expansion. Besides, the entire
Scorpius-Centaurus complex has a proper expansion with an angular
velocity $k_0=46\pm8$ km s$^{-1}$ kpc$^{-1}$ at the derived
parameters of the kinematic center $l_0=-40^\circ$ and $R_0=110$
pc. Based on this velocity, we estimated the characteristic
expansion time of the complex to be $T=21\pm4$ Myr. The proper
rotation velocity of the Scorpius-Centaurus complex is lower in
magnitude, is determined less reliably, and depends on the data
quality and on the parameters of the kinematic center.

Each of the LCC, UCL, and US groups and a number of such young
open clusters as $\beta$ Pic, TWA, Tuc/Hor, and Chamaeleon were
probably formed from a single parent cloud of hydrogen. The
structure of the $UV$-velocity distribution consisting of three
roughly parallel branches that we found leads us to conclude that
there were three periods of star formation separated by short time
intervals in the expanding Scorpius-Centaurus complex.

\bigskip
\subsection*{ACKNOWLEDGMENTS}

This work was supported by the Russian Foundation for Basic
Research (project no. 05--02--17047).

\newpage
{\large\bf REFERENCES}
\bigskip

1. V.A. Ambartsumian, Stellar Evolution and Astrophysics (AN Arm.
SSR, Yerevan, 1947) [in Russian].

2. V.A. Ambartsumian, Astron. Zh. 26, 3 (1949).

3. D. Barrado y Navascu\'es, J.R. Stauffer, et al., Astrophys. J.
520, L123 (1999).

4. A. Blaauw, Bull. Astron. Inst. Netherland 11, 414 (1952).

5. A. Blaauw, Ann. Rev. Astron. Astrophys. 2, 213 (1964).

6. A. Blaauw, The Physics of Star Formation and Early Stellar
Evolution, Ed. by C.J. Lada and N.D. Kylafis (Kluwer, Dordrecht,
1991).

7. V.V. Bobylev, Pis'ma Astron. Zh. 30, 185 (2004a) [Astron. Lett.
30, 159 (2004a)].

8. V.V. Bobylev, Pis'ma Astron. Zh. 30, 861 (2004b) [Astron. Lett.
30, 848 (2004b)].

9. V.V. Bobylev, Pis'ma Astron. Zh. 32, 906 (2006) [Astron. Lett.
32, 816 (2006)].

10. V.V. Bobylev, G.A. Gontcharov, and A.T. Bajkova, Astron. Zh.
83, 821 (2006) [Astron. Rep. 50, 733 (2006)].

11. D. Breitschwerdt and M.A. Avillez, Astron. Astrophys. 452, L1
(2006).

12. J.H.J. de Bruijne, Mon. Not. R. Astron. Soc. 310, 585 (1999).

13. W.J.B. Corradi, G.A.P. Franco, and J. Knude, Mon. Not. R.
Astron. Soc. 347, 1065 (2004).

14. D. Dravins, L. Lindegren, and S. Madsen, Astron. Astrophys.
348, 1040 (1999).

15. D. Fern\'andez, F. Figueras, and J. Torra, astro-ph/0611766
(2006).

16. B. Fuchs, D. Breitschwerdt, M.A. Avilez, et al., Mon. Not. R.
Astron. Soc. 373, 993 (2006).

17. E.J. de Geus, P.T. de Zeeuw, and J. Lub, Astron. Astrophys.
216, 44 (1989).

18. P.T. de Zeeuw, R. Hoogerwerf, J.H.J. de Bruijne, et al.,
Astron. J. 117, 354 (1999).

19. G.A. Gontcharov, Pis'ma Astron. Zh. 32, 844 (2006) [Astron.
Lett. 32, 759 (2006)].

20. The Hipparcos and Tycho Catalogues, ESA SP-1200, (1997).

21. E. Jilinski, S. Duflon, K. Cunha, et al., Astron.Astrophys.
448, 1001 (2006).

22. J.R.D. L\'epine and G. Duvert, Astron. Astrophys. 286, 60
(1994).

23. P.O. Lindblad, Astron. Astrophys. 363, 154 (2000).

24. P.O. Lindblad, P.O. Grape, K. Sandqvist, et al., Astron.
Astrophys. 24, 309 (1973).

25. L. Lindegren, S. Madsen, and D. Dravins, Astron. Astrophys.
356, 1119 (2000).

26. K.L. Luhman, Astrophys. J. 560, 287 (2001).

27. T.E. Lutz and D.H. Kelker, Publ. Astron. Soc. Pac. 85, 573
(1973).

28. S. Madsen, D. Dravins, and L. Lindegren, Astron. Astrophys.
381, 446 (2002).

29. V.V. Makarov, Astron. J. 126, 1996 (2003).

30. E.E. Mamajek, Astrophys. J. 634, 1385 (2005).

31. E.E. Mamajek and E.D. Feigelson, Astron. Soc. Pac. Conf. Ser.
244, 104 (2001).

32. E.E. Mamajek, W.A. Lawson, and E.D. Feigelson, Astrophys. J.
516, L77 (1999).

33. E.E. Mamajek, W.A. Lawson, and E.D. Feigelson, Astrophys. J.
544, 356 (2000).

34. E.E. Mamajek, M. Meyer, and J. Liebert, Astron. J. 124, 1670
(2002).

35. A.M. Mel'nik, A. K. Dambis, and A. S. Rastorguev, Pis'ma
Astron. Zh. 27, 611 (2001) [Astron. Lett. 27, 521 (2001)].

36. C.A. Murray, Vectorial Astrometry (Adam Hilger, Bristol, 1983;
Naukova Dumka, Kiev, 1986).

37. K.F. Ogorodnikov, Dynamics of Stellar Systems (Pergamon,
Oxford, 1965; Fizmatgiz, Moscow, 1965).

38. C.A. Olano, Astron.Astrophys. 112, 195 (1982).

39. A.E. Piskunov, N.V. Kharchenko, S. R\"oser, et al., Astron.
Astrophys. 445, 545 (2006).

40. M.E. Popova and A.V. Loktin, Pis'ma Astron. Zh. 31, 743 (2005)
[Astron. Lett. 31, 663 (2005)].

41. T. Preibisch and H. Zinnecker, Astron. J. 117, 2381 (1999).

42. M.J. Sartori, J.R.D. L\'epine, and W.S. Dias, Astron.
Astrophys. 404, 913 (2003).

43. G. Schaller, D. Schaerer, G. Meynet, et al., Astron. Astroph.,
Suppl. 96, 269 (1992).

44. J. Skuljan, J.B. Hearnshaw, and P.L. Cottrell, Mon. Not. R.
Astron. Soc. 308, 731 (1999).

45. I. Song, B. Zuckermann, and M.S. Bessel, Astrophys. J. 599,
342 (2003).

46. C.A.O. Torres, L. da Silva, G.R. Quast, et al., Astron. J.
120, 1410 (2000).

47. C.A.O. Torres, G.R. Quast, L. da Silva, et al., Astron.
Astrophys. 460, 695 (2006).

48. R. Wichmann, J.H.M.M. Schmitt, and S. Hubrig, Astron.
Astrophys. 399, 983 (2003).

49. B. Zuckerman, I. Song, and R.A. Webb, Astrophys. J. 559, 388
(2001).

\bigskip
Translated by V. Astakhov

%\end{document}
%%%%%%%%%%%%%%%%%%%%%%%%%%%%%%%%%%%%%%%%%%%%%%%%%%%%%%%%%%%%%%%%%%%%%%%%

\newpage
{
\begin{table}[t]                                                %% Table 1
\caption[]{\small\baselineskip=1.0ex\protect
 Data on Gould Belt stars HIP Cluster.

}
\begin{center}
\begin{tabular}{|c|c|c|c|c|c|c|c|c|c|c|c|}\hline
       &&&&&&&&&&\\
 HIP & Cluster & $V_r$ & $e_{V_r}$ & $i$ & Source &
 $P^{LCC}$ & $P^{UCL}$ & $P^{US}$ & $P^{Gould}$ & ... \\\hline
  1  &     2     &   3   &   4       &   5 &   6 &   7   &   8   &   9   & 10    & ... \\\hline
 50520 &  LCC & 15.8 & 0.7 & 2 & PCRV     & $0.06$ & $0.22$ & $0.00$ & $0.72$ & ... \\
 50847 &  LCC & 12.0 & 4.2 & 1 & PCRV     & $0.73$ & $0.00$ & $0.00$ & $0.27$ & ... \\
 53701 &  LCC & 15.7 & 4.3 & 1 & PCRV     & $0.00$ & $0.69$ & $0.00$ & $0.31$ & ... \\
 55425 &  LCC & 26.8 & 2.7 & 1 & Jilinski & $0.00$ & $0.00$ & $0.00$ & $1.00$ & ... \\
 56561 &  LCC & -1.1 & 2.5 & 2 & PCRV     & $0.00$ & $0.00$ & $0.00$ & $1.00$ & ... \\
 57851 &  LCC & 20.0 & 2.0 & 2 & P+Jil    & $0.00$ & $0.00$ & $0.50$ & $0.50$ & ... \\
 58867 & gLCC & -8.9 & 0.5 & 1 & Jilinski & $0.00$ & $0.00$ & $0.00$ & $1.00$ & ... \\
  ...  & ... & ... & ... & ... & ... & ... & ... & ... & ... & ... \\\hline
\end{tabular}
\end{center}
 {\small\protect
 Note: $e_\pi/\pi<0.2$, Table 1 is fully accessible in electronic form.}
\end{table}
}

%\end{document}
%%%%%%%%%%%
{
\begin{table}[t]                                                %% Table 2
\caption[]{\small\baselineskip=1.0ex\protect
 Kinematic parameters of the Scorpius-Centaurus association.

}
\begin{center}
\begin{tabular}{|c|r|r|r|r|r|r|r|r|r|}\hline
 &&&&&&&&&\\
 ¹ & $-U_0$ & $-V_0$ & $-W_0$ & $\omega_0$ & $\omega'_0$ & $k_0$ &
$k'_0$ & $\sigma_0$& $N_\star$ \\\hline
 1  & $-11.8_{(0.8)}$ & $-18.2_{(0.8)}$ & $-6.1_{(0.3)}$ & $20_{(8)}$ & $481_{(192)}$ & $40_{(8)}$ & $-171_{(192)}$ & $2.9$ & 134 \\
 2  & $-11.2_{(0.8)}$ & $-18.4_{(0.8)}$ & $-6.1_{(0.3)}$ & $29_{(8)}$ & $421_{(193)}$ & $46_{(8)}$ & $  17_{(193)}$ & $2.9$ & 134 \\
 3  & $-11.2_{(0.8)}$ & $-18.6_{(0.8)}$ & $-6.1_{(0.3)}$ & $38_{(8)}$ & $356_{(195)}$ & $34_{(8)}$ & $  14_{(195)}$ & $2.9$ & 134 \\\hline
 2a & $ -8.0_{(1.3)}$ & $- 8.3_{(1.3)}$ & $-6.1_{(0.1)}$ &$-56_{(8)}$ & $125_{(136)}$ & $41_{(8)}$ & $ 135_{(136)}$ &$1.7$ & 279 \\
 2b & $ -8.2_{(1.2)}$ & $- 9.3_{(1.2)}$ & $-6.5_{(0.2)}$ &$-40_{(11)}$ & $232_{(200)}$ & $39_{(12)}$ & $ 347_{(183)}$ &$1.7$ & 279 \\\hline
\end{tabular}
\end{center}
 {\small\protect
 Note:
 $U_0, V_0, W_0$, and $\sigma_0$ are in km s$^{-1}$,
 $\omega_0$ and
 $k_0$ are in km s$^{-1}$ kpc$^{-1}$,
 $\omega'_0$ and
 $k'_0$ are in km s$^{-1}$ kpc$^{-2}$,
 $N_\star$ is
the number of stars used, all quantities were determined with the
parameters of the center $l_0=-40^\circ$ and $R_0=110$ pc,
$\sigma_0$ is the error of a unit weight found when solving the
system of equations (1)--(3).}
 \end{table}
}

%%%%%%%%%%%%%%%%%%%%%
\newpage
%%%%%%%%%%%%%%%%%%%%%%%%%%%%%%%%%%%%%%%% FIG.1:
\begin{figure}[t]
{
\begin{center}
  \includegraphics[width=100mm]{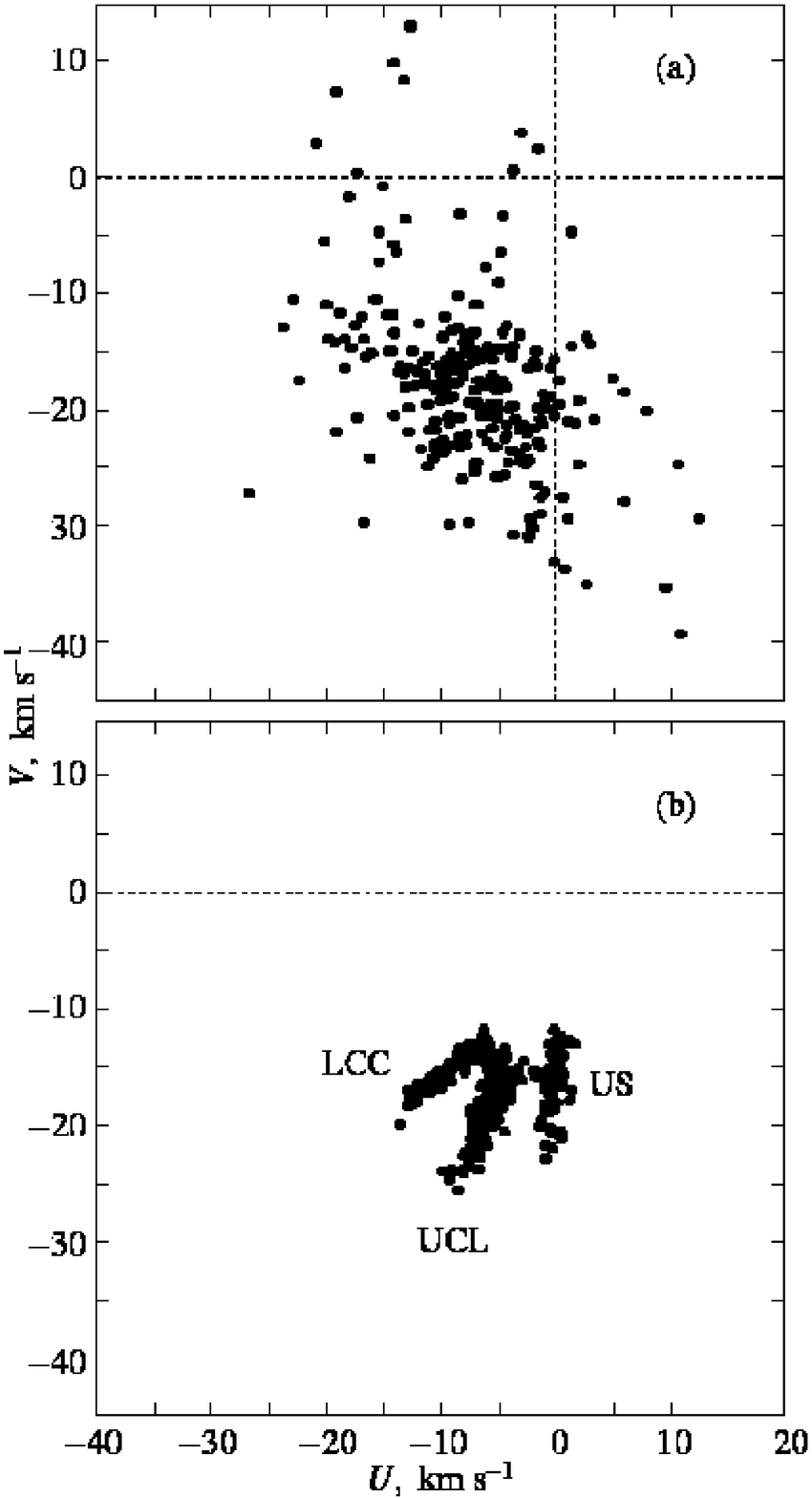}
\end{center}
} Fig. 1. (a) $UV$ velocities of 255 Gould Belt stars (``real''
data); (b) $UV$ velocities of 380 Scorpius-Centaurus stars whose
radial velocities were taken from the catalog by Madsen et al.
(2002). The stellar velocities were corrected for the Galactic
rotation and are given relative to the Sun.
\end{figure}

\newpage
%%%%%%%%%%%%%%%%%%%%%%%%%%%%%%%%%%%%%%%% FIG.2:
\begin{figure}[t]
{
\begin{center}
  \includegraphics[width=100mm]{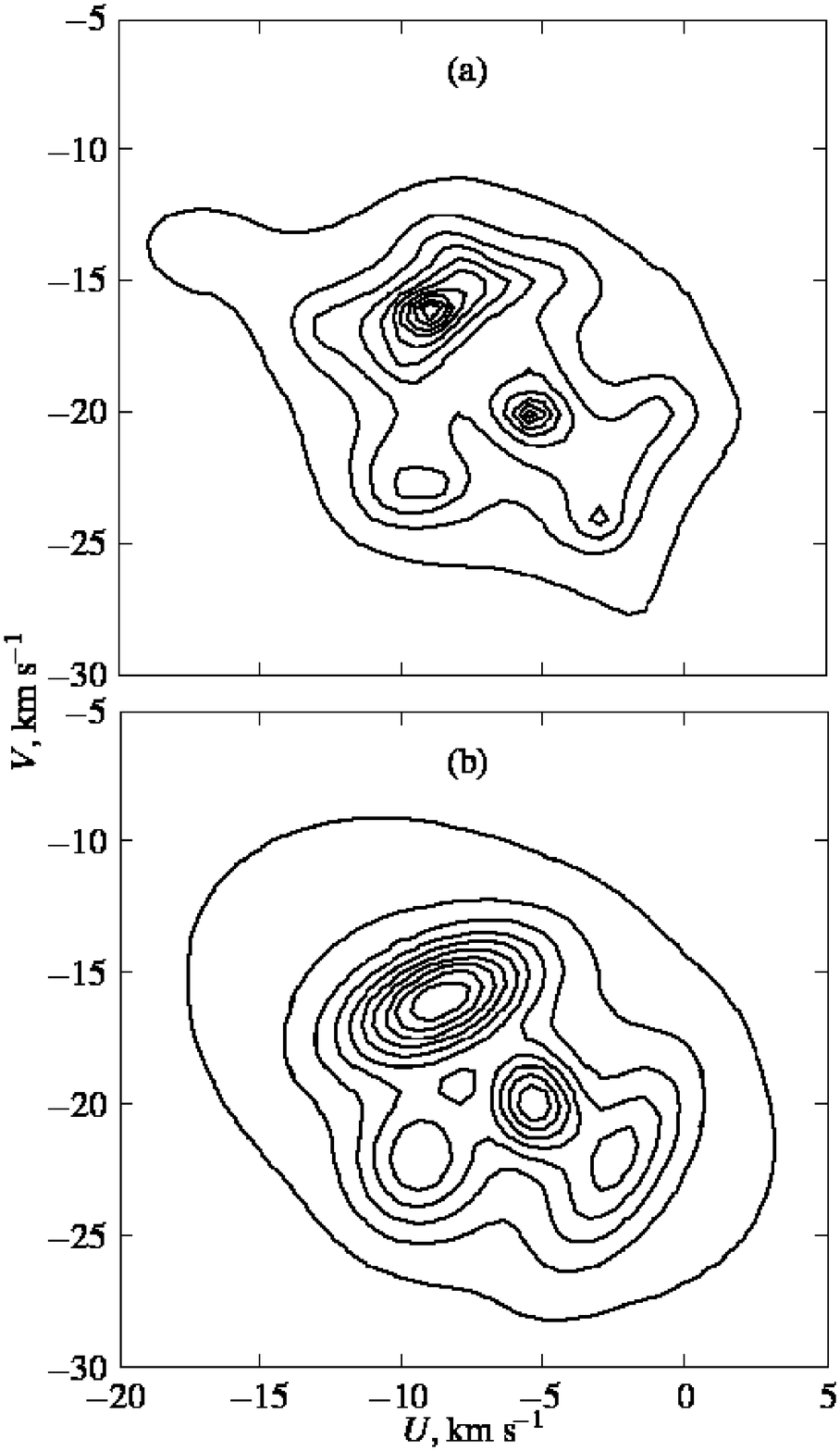}
\end{center}
}
Fig. 2. (a) Probability density distribution obtained by the
adaptive kernel method from a given discrete distribution of
stellar UV velocities; (b) distribution obtained by approximating
the UV velocity distributions of stars of individual fractions by
individual Gaussians.
\end{figure}

\newpage
%%%%%%%%%%%%%%%%%%%%%%%%%%%%%%%%%%%%%%%% FIG.3:
\begin{figure}[t]
{
\begin{center}
  \includegraphics[width=110mm]{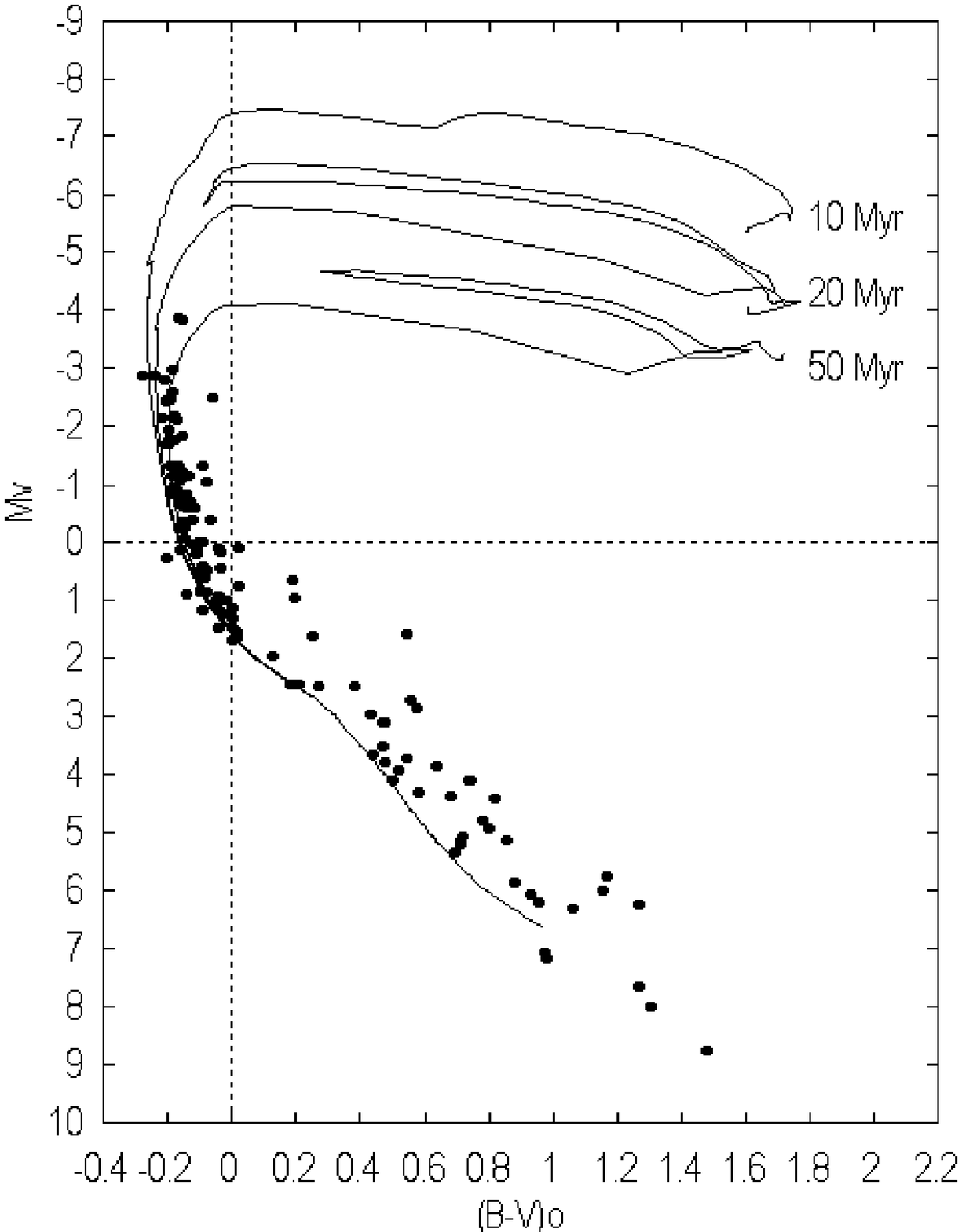}
\end{center}
} Fig. 3. Color-absolute magnitude diagram for 134
Scorpius-Centaurus stars that satisfy the condition
$e_\pi/\pi<0.15$.
\end{figure}

\newpage
%%%%%%%%%%%%%%%%%%%%%%%%%%%%%%%%%%%%%%%% FIG.4:
\begin{figure}[t]
{
\begin{center}
  \includegraphics[width=110mm]{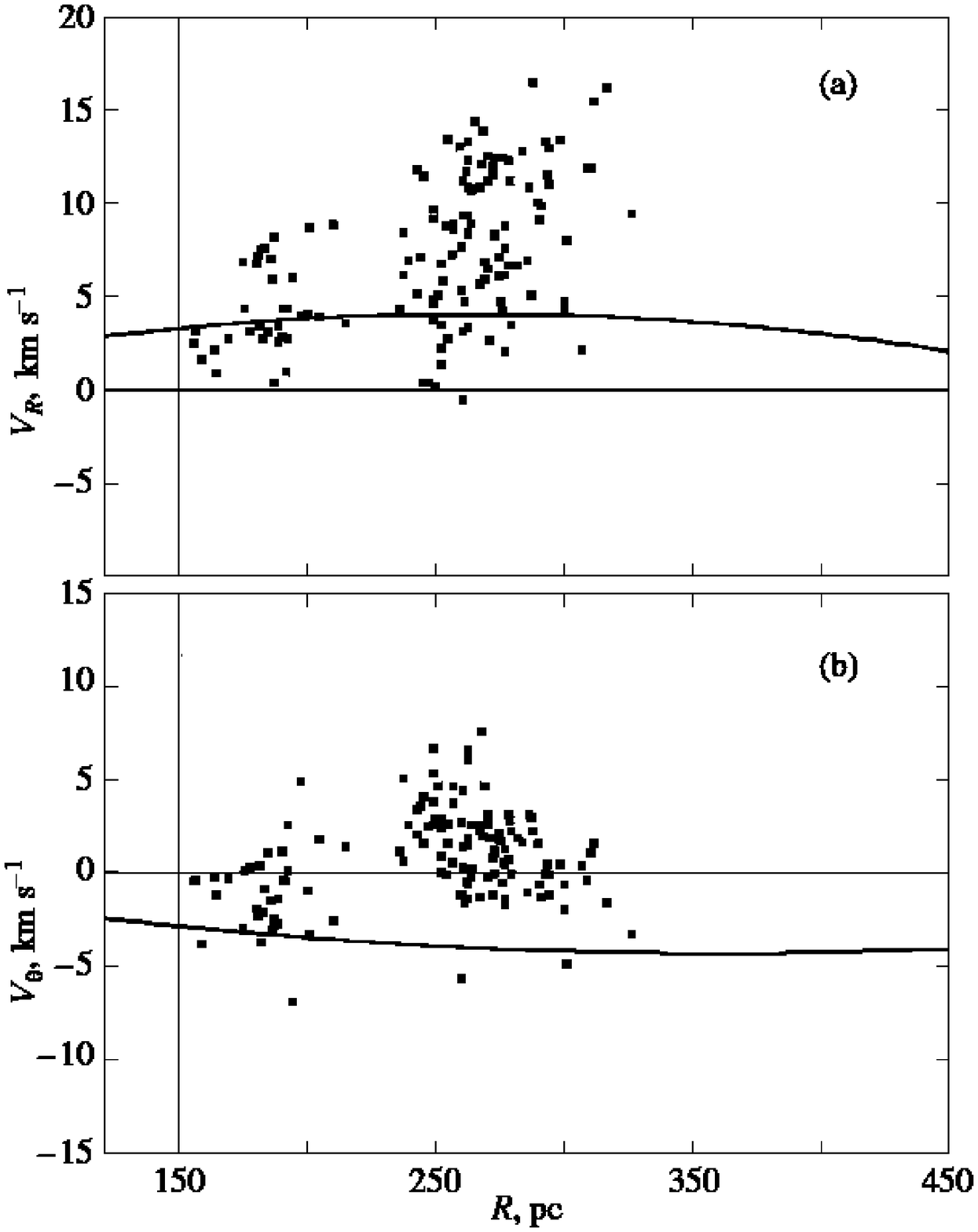}
\end{center}
} Fig. 4. Expansion, $V_R$ (a), and rotation, $V_\theta$ (b),
velocities versus distance $R$ from the kinematic center of the
Gould Belt calculated at $l_0=128^\circ$ and $R_0=150$ pc; the
vertical line marks $R_0$.
\end{figure}

\newpage
%%%%%%%%%%%%%%%%%%%%%%%%%%%%%%%%%%%%%%%% FIG.5:
\begin{figure}[t]
{
\begin{center}
  \includegraphics[width=120mm]{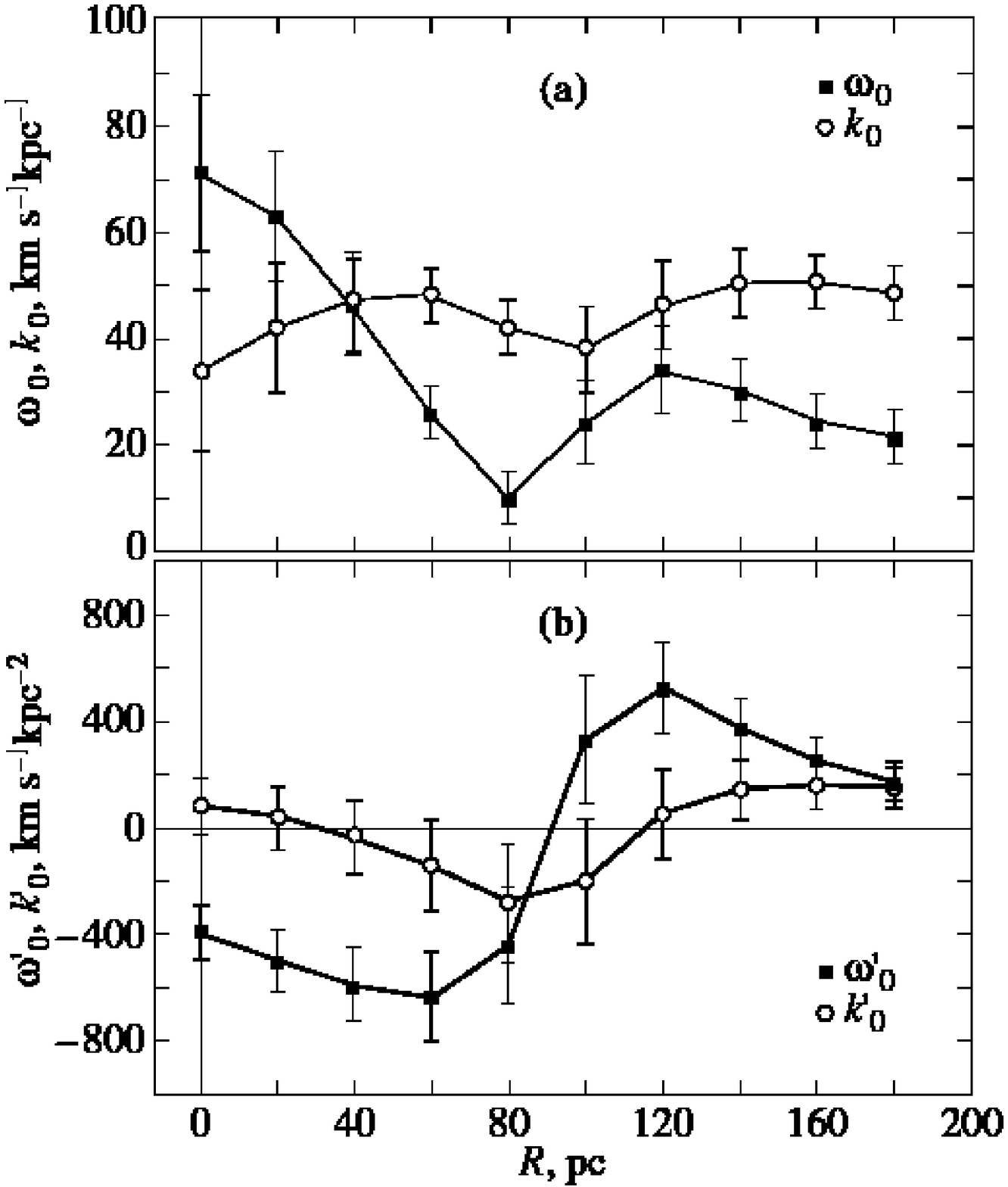}
\end{center}
} Fig. 5. (a) Parameters of the angular velocities of rotation and
expansion for the Scorpius-Centaurus association and (b) the
corresponding derivatives versus distance from the kinematic
center of the association calculated at $l_0=-45^\circ$.
\end{figure}

\newpage
%%%%%%%%%%%%%%%%%%%%%%%%%%%%%%%%%%%%%%%% FIG.6:
\begin{figure}[t]
{
\begin{center}
  \includegraphics[width=110mm]{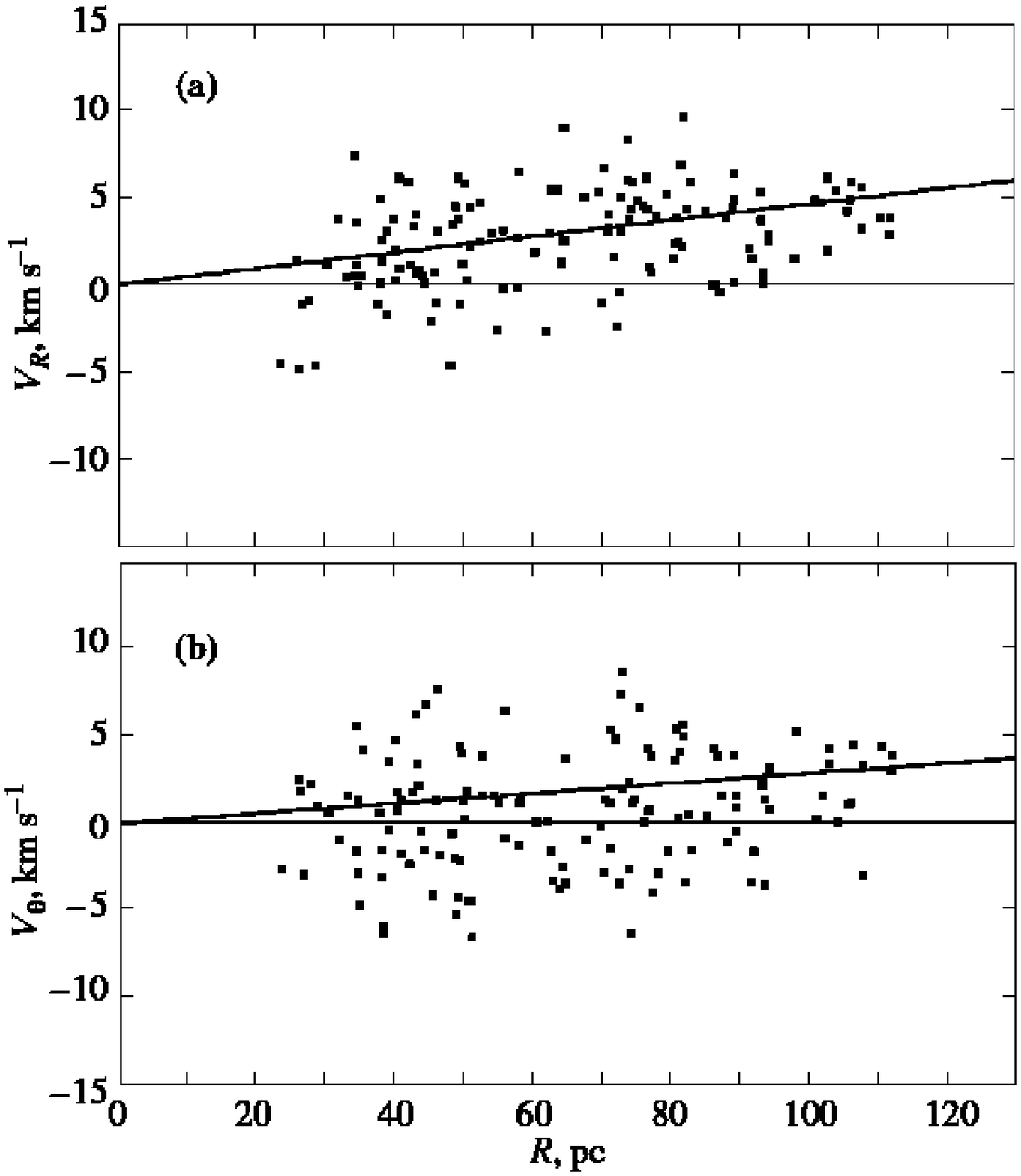}
\end{center}
} Fig. 6. Expansion, $V_R$ (a), and rotation, $V_\theta$ (b),
velocities versus distance R from the kinematic center of the
Scorpius-Centaurus association calculated at
 $l_0=-40^\circ$ and
 $R_0=110$ pc; the vertical line marks $R_0$.
\end{figure}

\newpage
%%%%%%%%%%%%%%%%%%%%%%%%%%%%%%%%%%%%%%%% FIG.7:
\begin{figure}[t]
{
\begin{center}
  \includegraphics[width=110mm]{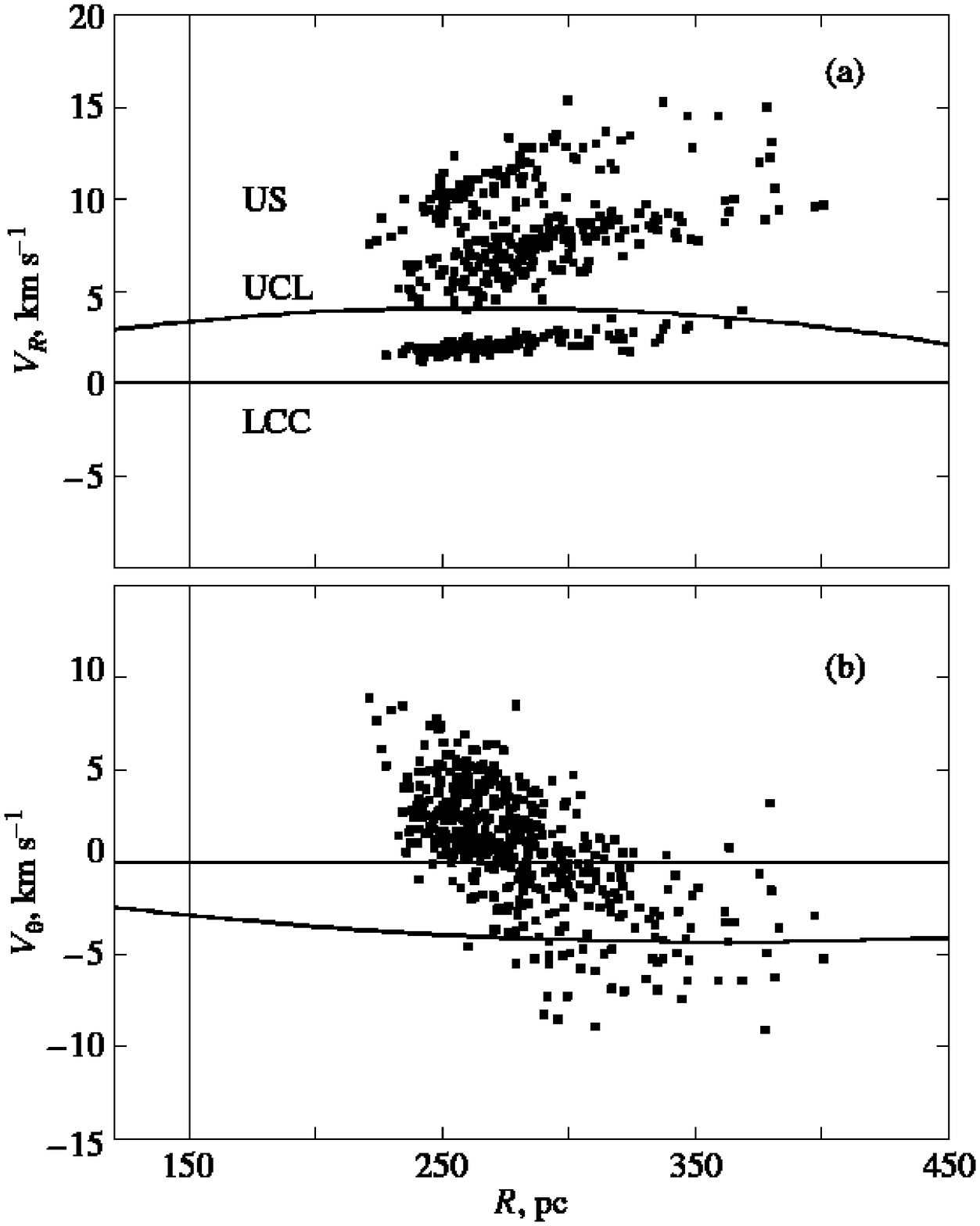}
\end{center}
}
Fig. 7. Expansion, $V_R$ (a), and rotation, $V_\theta$ (b),
velocities of 489 Scorpius-Centaurus stars whose radial velocities
were taken from the catalog by Madsen et al. (2002) versus
distance $R$ from the kinematic center of the Gould Belt
calculated at $l_0=128^\circ$ and $R_0=150$ pc; the vertical line
marks $R_0$.
\end{figure}

\end {document}